\documentclass{nws}
\usepackage{lastpage}

\title[Gender Differences in Social Networks]
      {Gender Differences in Communication Behaviors, Spatial Proximity Patterns, and Mobility Habits}

 \author[Yang Yang et al.]{
    YANG YANG \textsuperscript{1,2},
    OMAR LIZARDO \textsuperscript{2,3},
    DONG WANG \textsuperscript{1,2},
    YUXIAO DONG \textsuperscript{1,2},
    AARON D. STRIEGEL \textsuperscript{1,2},
    DAVID HACHEN \textsuperscript{2,3},
    and NITESH V. CHAWLA \textsuperscript{1,2} {\footnote{Corresponding Author}} \\ 
    \textsuperscript{1} Department of Computer Science and Engineering, University of Notre Dame \\ 
    \textsuperscript{2} Interdisciplinary Center for Network Science and Applications\\
    \textsuperscript{3} Department of Sociology, University of Notre Dame \email{nchawla@nd.edu}}

\begin{document}

\label{firstpage}

\maketitle
\begin{abstract}
The existence of gender differences in the structure and composition of social networks is a well established finding in the social and behavioral sciences, but researchers continue to debate whether structural, dispositional, or life course factors are the primary driver of these differences. In this paper we extend work on gender differences in social networks to patterns of interaction, propinquity, and connectivity captured via a \emph{social sensing} platform comprised of an ensemble of individuals' phone calls, text messaging, face-to-face interactions, and traces of their mobility activities. We attempt to isolate dispositional from other factors by focusing on a relatively homogeneous population on a relatively closed setting at the same stage in the life course. Analysis across three different networks along with mobility data reveals systematic gender differences in terms of communicative, distributional, mobility, and spatial proximity tendencies. Most importantly, some patterns observed in the communication network (e.g. stronger same-gender preference for women) are found to be reversed in the spatial proximity context, with men displaying a greater tendency to spend time in a narrower (and thus more predictable) range of spaces with same-gender peers than women. These findings provide robust and novel evidence for the powerful effect of gender in structuring behavior across a wide range of communication and mobility behaviors, validating and extending recent work pointing to differences of socio-cultural and evolutionary origin in the styles of sociability and interaction characteristic of men and women.

\end{abstract}


\section{Introduction}
Gender is one of the most important individual-level attributes known to structure patterns of social connectivity in human social networks \cite{ridgeway1999, ridgeway2006}. In Sociology, scholars have documented the organization of interaction according to gender lines  across private, formal, informal, and public spheres of association \cite{ridgeway2004, mcpherson2001, munch1997, mcpherson1986, ibarra1997, hagan1998, smith1993}. Given this, it is not surprising to find that beginning with some of the earliest attempts to introduce a network perspective into the social sciences \cite[e.g.]{moreno1953, bott1955}, researchers have found consistent differences in the relative structure, composition, and relational quality of men and women's social networks \cite{fischer1982, marsden1987, coates1987, moore1990, dunbar1995, mcpherson2006}.

For instance, women have been found to include more kin as contacts with whom they discuss ``important matters'' or get social and emotional support than men \cite{fischer1983, marsden1987, coates1987, moore1990, dunbar1995}; to be more likely to generate a larger number of ``strong ties'' (contacts with whom the feel emotionally close) in free recall of friends and acquaintances \cite{fischer1982, marsden1987}; to be less likely to include co-workers in their core discussion networks \cite{fischer1983, marsden1987, moore1990}; and to be more likely to interact with members of their personal in private or informal settings rather than in public settings or formal organizations \cite{mcpherson1986, coates1987, marsden1987, mcpherson1986}.

In addition, the observation that friendship choices in social networks display a strong pattern of ``gender-segregation'' (also referred to as ``gender homophily'' or ``assortative mixing by gender'') is one of the most well-replicated and long-standing empirical generalizations in the social and behavioral sciences \cite{maccoby1988, benenson1990, smith1993, mcpherson2001, mehta2009}. The basic observation here is that same-gender preferences in friendship choices emerges very early in the life-course (at about 3 to 4 years of age), strengthens during adolescence and remains fairly strong through young adulthood, middle and old age \cite{mehta2009}. The other key finding in this literature is that women tend to display significantly stronger same-gender preference in friendships especially for close friends \cite{mehta2009}, although this pattern waxes and wanes depending on the stage in the life-course \cite{fischer1983, palchykov2012}.

Taken together, research on gender differences in the structure and composition of social networks along with work examining gender differences in communication propensity, partner preferences, and relationship maintenance style have the potential to generate diffusion and wide-randing effects in the macro-level structure of human social networks. In this paper we contribute to the literature on the origins of compositional differences in men and women's social networks. We do this both by replicating long-standing findings in the literature (e.g. stronger same-gender preferences for women in communication networks) and by extending the gender-differences paradigm to a new set of social network metrics (e.g. distributional tendencies across alters and gender categories in communication and proximity networks) and behavioral indicators (e.g. predictability of spatial mobility patterns). 


\section{Theoretical Approaches to Studying Gender Differences in Social Networks}
Social network theorists have proposed two basic approaches to understanding the origins of gender differences in the structure and composition of social networks: \emph{dispositional} and \emph{structural} \cite{moore1990, smith1993}. Dispositional explanations for gender differences in social network structure and composition point to enduring, habitual behavioral tendencies (e.g. towards reciprocation, transitivity, frequency of communication, and so on) that make women's relationship building and maintenance style distinctive from that of men \cite{fischer1983}. Dispositional theorists point to the origins of these enduring behavioral tendencies in a combination of cultural, socialization, and personality-level influences. Structural theorists on the other hand explain differences in the structure and composition of men and women's networks by pointing to the different positions occupied by men and women in the social system, especially when these different positions entail differential access and opportunity to generate social connections of a given type (e.g. weak ties versus strong ties). 

A third approach, the \emph{life cycle} perspective, attempts to mediate between dispositional and structural perspectives by noting that both factors may be at play but that whether one or the other is most prevalent depends on an individual's age \cite{kalmijn2002}. In this respect, life-cycle theorists point to the possibility that gender differences in personal network structure and composition may be due more to the stage of the life-cycle in which persons find themselves. From this perspective, family formation, entry into (or exit from) the workforce, the advent of children, and the unequal distribution of family responsibilities generates tendencies for women to draw on kin ties and strong ties for social and emotional support \cite{fischer1983, munch1997, palchykov2012}. 

Previous work attempting to isolate dispositional differences in men and women's approach to relationship formation and maintenance has had trouble separating these from structural, and life-cycle influences. This line of research has suffered from two major drawbacks: First, there has been an undue focus on static qualities of personal networks obtained from free-recall of personal contacts obtained via the name generator method \cite{stehle2013}. Second, average gender differences are computed for groups who are either at different stages in the life cycle or are located different structural positions, thus combining information from persons in both distinct positions in the structure and at different stages in the life-cycle. 

Reliance on personal networks constructed from free-recall of personal contacts elicited by name generator prompts have been subject to criticisms on various methodological grounds \cite{brewer2000, marsden2003, marin2004}. For our purposes, the key limitation of this approach when it comes to drawing inferences related to the dispositional origins gender differences in social networks is that we cannot separate gender differences connected to cognitive and emotional biases in recall and naming of contacts from dispositional differences connected to communication \emph{behavior}. For instance, the fact that women tend to name more kin and ``close" contacts than men, may be due to the fact that both kin and emotionally close contact are more cognitively and affectively salient for women or that they women are in fact more likely to sustain ties that require a high level of behavioral engagement. Self report of the number of ties to which the persons feels subjectively close, however, may not tell us much about dispositional differences in behaviors relevance to network formation and relationship maintenance. In what follows we deal with this issue by focusing our attention on \emph{behavioral} differences in the social networking behavior of men and women obtained via non-obtrusive data collection strategies tied to a remote social-sensing platform keyed to automatically capture social interactions between individuals \cite{striegel2013}.

Reliance on groups that are heterogeneous on both life-cycle stage and structural location on the other hand, leads to problems in isolating behavioral differences relevant to the structure and composition in men and women's social networks that come from differences in opportunity structure, gender inequalities in the domestic division of labor, childcare, and workforce participation \cite{munch1997, smith1993, moore1990}. For instance, the over-representation of kin and the stronger same-gender preference patterns in women's networks may be due to dispositional factors (actual preference for social contact with kin and same-gender alters) or to the fact that women continue to be more likely to do the bulk of household related and childcare related activities \cite{fuwa2004}, in settings where they are more likely to come into contact with other women and draw on close relatives for help and support \cite{munch1997, smith1993}. 

In what follows, we isolate dispositional from structural and life course factors by focusing our analysis on a population that is largely homogeneous in terms of life-course stage (e.g. unmarried, childless, young adults between the ages of 17-19 making the transition to college), located in a setting that provides most persons with a similar opportunity structure for network formation (a self-contained, majority-residential campus, located in a mid-sized town). This is an ideal time in the life-course to study dispositional differences in network-related behaviours as the life-course events associated with gender-based structural differences in the composition and structure of social networks such as marriage, labor-force entry and childbearing \cite{fischer1983, moore1990, munch1997}, have yet to take effect. 

\section{Previous work on Gender Differences in Social Networks Using Unobtrusive Data Collection Strategies}
While not extensive, an emerging  literature using unobtrusive data collection strategies points to possibility of important dispositional differences in the relational behavior of men and women especially when it comes to revealed preferences and usage patterns of remote telecommunication media, such as phone calls or text messaging. 

\subsection{Mobile Communication Networks}
Several studies on gender differences in relational and network behaviors relies on mobile communication network (MCNs) data constructed from large data sets (usually containing millions of users) of mobile phone call data records. For instance, a study of ``temporal motifs'' finds that all-female subgraphs of size three are over-represented in these types of networks \cite{kovanen2013}, suggesting that women are (a) more likely to initiate communication events and direct them at other women (``out-star'' temporal motif), (b) more likely to be the recipient of communication events from other women (``in-star'' temporal motif), (c) induce their female contacts to generate subsequent communication events of their own post-contact (``causal chain'' temporal motif), and (d) be the recipient of a communication event from another woman after directing a communication event to a seemingly unrelated contact (``non-causal chain'' temporal motif). 

Additional MCN-based studies have been able to replicate the classic observation of strong same-gender preference patterns for each user's strongest dyadic connections in the context of mobile phone communication data \cite{palchykov2012}. Taking a life-course perspective, this work shows that same-gender preference for the strongest link is pervasive and strong at all ages, but that (a) it is stronger for women, and (b) the strength of the gender difference depends on life course stage \cite{palchykov2012}. Finally, other MCN-based work shows that women are more likely to engage in cross-generation communicative interactions than men, tend to call more frequently and spend more time on the phone, and display a stronger preference for same-gender communication \cite{stoica2010, dong2014}. This work replicates stylized facts regarding women's propensity towards greater sociality and kin connectivity that had been uncovered in previous social network research using free-recall name-generators \cite{fischer1983, marsden1987, moore1990}.

\subsection{Online and Social Media Networks}
Research dealing with gender differences in the context of online social networks sites (SNSs) replicates the long standing finding of stronger assortative mixing preferences for women. This same research, however shows that cross-gender connections are more common for users with a larger circle of friends \cite{volkovich2014}. While not sharing all of the features of social interactions in natural settings, recent work reveals strong gender-specific effects in relational behavior in SNSs dedicated to online gaming \cite{szell2013}. This work finds that women tend to be more risk averse, while demonstrating stronger same-gender preferences (assortative mixing) than men in this setting. Related work finds that these type of settings elicit preferential behavior directed towards women, including a larger volume of incoming communication \cite{griffiths2004}. Using data from Facebook profiles for a population very similar to the one under consideration here (the Fall of 2005 incoming Freshman cohort at a ``diverse private college in the Northeast U.S."), Harvard researchers found that women tended to have larger, more active, and more diverse Facebook networks than men \cite{lewis2008}.

\subsection{Spatial Proximity Networks}
Recent work has begun to tackle issues of gender differences in network related behaviors by constructing spatial proximity networks (SPNs) from wearable sensing badges \cite{paradiso2010}. These sensors, worn continuously by study participants, are designed to capture each individual location signature patterns, allowing researchers to study proximity behaviors. This work shows that generalizations obtained from communication network analysis (e.g. mobile phones and social media), such as the stronger preference for same-gender contact displayed by women, may not be completely generalizable to spatial proximity networks.

For instance, in a study of contact proximity among high school students, it was found that while female students seemed to display little same-gender preference (in fact spending most of their time around other-gender students) in proximity behaviors, male students were much more likely to spend the majority of their time in close proximity to other male students. These differences were partially attenuated once the marginal distributions of students of each gender were accounted \cite{fournet2014}. A recent study of of primary school children (aged 6 - 12) using wearable sensor methodology finds similar results. While same-gender preference in spatial proximity behaviors is found for boys and girls of of all ages, boys tend to display a stronger pattern of same gender spatial proximity preference and this effect becomes larger as boys get older \cite{stehle2013}. In all, emerging results on patterns of interactive gender-segregation in spatial proximity networks using unobtrusive methods contrasts sharply with the recurrent pattern of stronger same-gender preference for women found in the work communication-based networks considered above \cite{stoica2010, palchykov2012}. 

\section{Our Work}
Social sensing has emerged as a new paradigm of crowdsourcing the data collection tasks to average individuals~\cite{wang2015social}. Our work leverages high quality information from a social sensing platform to provide a wide-ranging exploration of dispositional differences in men and women's relationship building and relationship maintenance style that is relatively (but not completely) unconfounded by life-course and structural influences. We move beyond previous work by studying patterns of communication among young adults making the transition to college in a natural setting in a context in which they are in the process of ``turning over'' their previous (high school) network and forming new relationships \cite{oswald2003, kane2011}. 

This unique dataset \cite{meng2014} allow us to consider a wider range of network-relevant communication behaviors and network formation tendencies than previous work on the subject. In particular we seek to move beyond the emphasis on gender differences in static characteristics of social networks such as assortative mixing and reciprocity \cite{rivera2010, szell2013}. We focus on gender-specific difference in managing and maintaining interactions via three communication channels (phone call, text messaging, and via spatial proximity). We further examine how individuals sustain or modify their behaviors in these three types of communication channels in their turn. Gender differences on each criterion behavior are assessed via the construction of several null models, in which we compare observed gender patterns to what we would expect by chance \cite{kovanen2013,yang2014}. 

In keeping with the argument outlined above, we focus on dispositional differences that can be tied most directly to behavior (and not issues of the cognitive or emotional salience of personal contacts). In particular we focus on broad behavioral patterns that may be tied to dispositional differences between men and women. Some of these patterns have been addressed in previous work (e.g. same-gender preference in communication and proximity networks), while others are analyzed for the first time here (e.g. gender differences in distributional tendencies and spatial mobility diversity).

\subsection{Gender Differences in Social Networks}
We address the question of the extent to which we can observe a pattern of same-gender preference across communication (voice call and text messaging) and spatial proximity networks. We also address the question of whether there are gender differences in the strength and direction of the same-gender preference effect across communication and proximity networks. Our review of previous studies suggests that we should find strong patterns of gender based homophily across these networks. However, given the different pattern of results obtained in the communication network and spatial proximity studies \cite{fournet2014, stehle2013}, we anticipate finding stronger same-gender preference patterns for women in communication networks, and stronger same-gender preference patterns for men in spatial proximity networks. In addition, we address the question of whether there exist gender differences in the \emph{coupling} of communication and proximity. To address this issue, we examine the extent to which we can observe: (a) increases in dyadic proximity behaviors following connectivity in the communication networks, and (b) the extent to which the strength of this effect differs for men and women. 
    
\subsection{Gender Differences in Sociability, Popularity and Reciprocity}
Previous work on MCNs (mobile communication networks) and SNSs (social networks sites) has shown strong differences between men and women in both their tendencies to direct and receive communications from others, with women being more likely to engage in and receive communications than men. We ascertain whether we can observe similar differences in these two behavioral outomes in our two directed (voice call and text messaging) networks. Reciprocity is a key construct in social network theory \cite{hallinan1979}, and an important behavioral marker in social interactions \cite{gouldner1960}. Here we address the question of whether there exist gender differences in the tendency to reciprocate communicative interactions. 
    
\subsection{Tendencies Towards Equitable Distribution of Communications}
While previous work has examined the general tendencies of individuals to distribute their communications unequally across contacts \cite{miritello2013}, the question of whether men and women differ on this crucial behavioral trait has not been examined before. This is largely because of the unavailability of data containing fine-grained records of communication frequencies across contacts \cite{eagle2010}. We introduce a quantitative metric of an individual's ability to distribute their communications equitably (or inequitably) across their contacts and ascertain whether men and women are different along this metric. We extend this work by examining the extent to which men and women differ not only in their distribution across contacts, but also in their distribution tendencies across \emph{types of contacts} (classified by gender). In addition, we also extend this framework to examine gender differences in distributive tendencies regarding the \emph{time} spend with other persons in the spatial proximity network.
    
\subsection{Gender Differences in Location Diversity}
Different relationships are tied to different spatial locations \cite{doreian2012}. In addition, research in the emerging field of human mobility studies reveals surprising levels of predictability in human mobility behavior \cite{song2010}. If the structure of men and women's social networks differ, this may be tied to dispositional differences in the way that men and women distribute their time across spatial locations \cite{mehta2009}. However, this is a question that has yet to be addressed in social network and mobility studies. We have at our disposal detailed records the traces of students in 54 locations on campus, which allows us to explore gender-specific difference patterns of mobility (Figure~\ref{fig_mobility_flow}). We examine three questions: Are men and women different in the relative \emph{diversity} (e.g. predictability or lack thereof) of their mobility patterns? Are same-gender dyads more similar in mobility patterns than cross-gender dyads? Do mobility patterns change after persons become connected in the communication network?

\subsection{Paper Organization}
The rest of the paper is organized as follows, section~\ref{sec_data} we introduce and describe our main data source. We follow with a detailed description of preliminaries in section~\ref{sec_prelim}. In section~\ref{sec_analysis} we report the key dispositional gender differences that emerge in our analysis. We conclude by outlining the implications of our results for future work on gender differences in social networks in section~\ref{sec_discussion}.

\section{Data}
\label{sec_data}
The ND mobility data \cite{meng2014} is drawn from the University of Notre Dame's NetSense smartphone study \cite{striegel2013}. This study was launched in August of 2011 with the goal of monitoring the communication and relationship formation behavior of 200 freshmen entering the university in the Fall of 2011. The primary data collection strategy of social interactions was via unobtrusive social sensing. Study participants were equipped with smartphones modified to track, the location, calls and texts made and received (but not the content of their communications) over a two-year period using a monitoring app employed on each phone. This app logged and then transmitted to a secure database a Call Data Record (CDR) for each communication event.  

Each CDR contains the phone numbers of the sender and receiver along with a timestamp indicating when the event occurred. There are two broad categories of communication types among these students: digital communications (text messages and phone calls) and face-to-face interactions (proximity observed via Bluetooth). Additionally we also have location traces of each student. There are in total 54 locations labelled in this dataset; among them 18 locations are for study usage (i.e., classrooms and library), 27 of them are residence halls, and rest of them are church, football stadium, dining hall and etc. All of data described above are collected between August of 2011 and December of 2012. The data collection spans across three semesters: 1) 2011-08 to 2011-12; 2) 2012-01 to 2012-05; 3) 2012-08 to 2012-12. We consider the following four categories of data.

\begin{figure*}[ht]
	\centering
	\centerline{\includegraphics[width=3in]{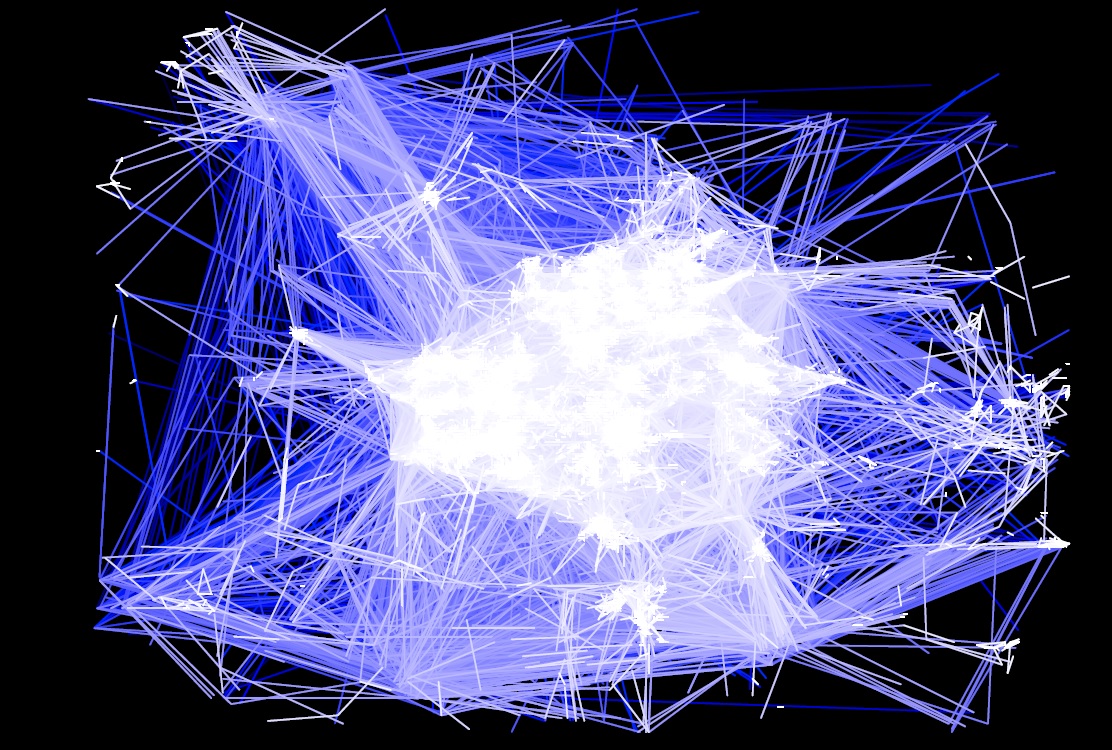}}
	\caption{Mobility Flows on Campus. The color ranges from blue to white. White color indicates that the density of flux is high, while blue means that the flux is rare.}
\label{fig_mobility_flow}
\end{figure*}

\subsection{Voice Call Network}
We built the voice call network from the CDR log. In the voice call networks vertices are individual students and an edge exists between individual $u$ and an individual $v$ if there is CDR showing that $u$ initiated a voice call with $v$. The voice call network is both \emph{directed} and \emph{weighted}.

\subsection{Text Message Network} 
In the same way, we built the text message network from the CDR log. In the text message network vertices are individual students and an edge exists between individual $u$ and an individual $v$ if there is CDR  showing that $u$ sent a text message to individual $v$. Like the voice call network, the text message network is both \emph{directed} and \emph{weighted}.

\subsection{Spatial Proximity Network}
In the proximity network, vertices are individual students and an edge exists between individual $u$ and an individual $v$ if bluetooth logs show them to be proximate to one another at that time. The weight of the $(u,v)$ edge is given by the total duration that they have been proximate as the weight of the link $(u,v)$. Given the inherent non-directionality of the proximity relation, this network is \emph{undirected} and \emph{weighted}.\footnote{We only consider proximity network in three semesters, due to the fact that there are rare proximity interactions when students are on vacations. Thus our validations and experiments are conducted on semester 1 (2011-08 to 2011-12), semester 2 (2012-01 to 2012-05), and semester 3 (2012-08 to 2012-12) proximity networks.}

\subsection{Location Traces}
The GPS device in each cellphone also records the location traces of each student between August of 2011 and December of 2012. If an individual student is at the location $L$, then the monitor software will send this information with the current timestamp to a server. Thus we have the location trace for each student, which allows us to analyze their mobility behaviors (Figure~\ref{fig_mobility_flow}). We divided campus location into 54 distinct spatial regions. Among these, 18 are dedicate for schooling and study activities (e.g., classrooms and library), 27 are (same-gender only) residence halls, and the rest consist of such function and activity specialized locations as the Basilica (Church), the football stadium, the dining halls (two major ones ``North'' and ``South''), and so on. In this way we are able construct the location visiting signature $T_{L} (u)$ for each individual $u$. We can then compute the similarity of location visiting patterns between any pair of students. Additionally a student $u$'s location signature $T_{L} (u)$ allows us to calculate the extent to which he or she spends time at a small number of locations or spreads his or her time evenly across locations (see Equations~\ref{eq:location} and~\ref{eq:locdiv} below). 


\section{Network Metrics}
\label{sec_prelim}

One main goal of this study is to determine the existence of dispositional differences between men and women in communication behavior, such as the distribution of communicative behaviors across alters (among others). In this section we present and formalize several behavioral metrics aimed at capturing gender differences in relationship formation and relationship maintenance behavior across our four different sources of relational data, namely phone call, text messaging, face-to-face interaction, and mobility traces. 

\subsection{Topological Diversity}
\begin{figure*}[h]
\begin{center}
\centerline{\includegraphics[width = 3.2in]{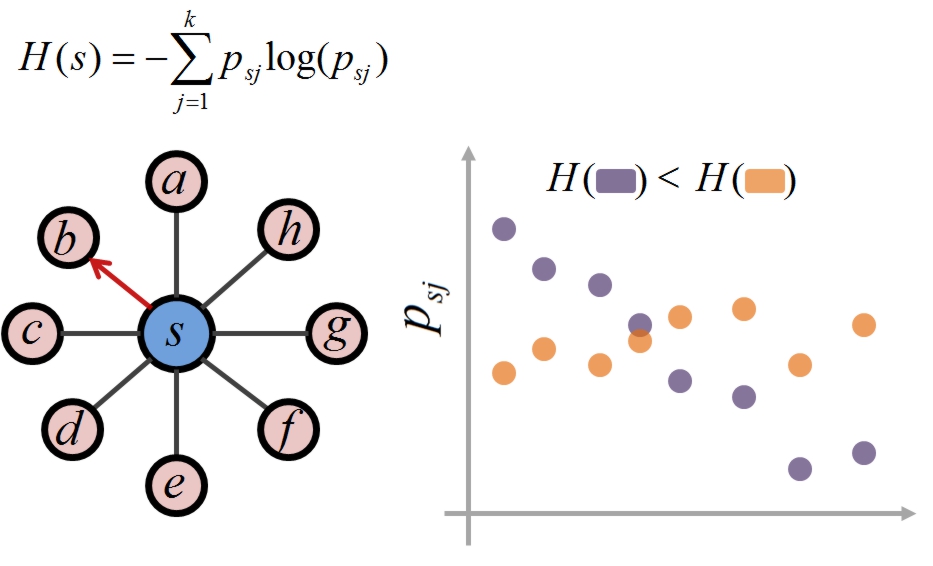}}
\caption{Ego Network diversity for a hypothetical individual. If an individual's communications are distributed evenly across different alters (brown), then the diversity value is high; conversely, if the bulk of communications go to a small number of alters (blue), the diversity value is small.}
\label{fig_diversity}
\end{center}
\end{figure*}

We use Shannon's entropy \cite{eagle2010,shannon1949} to capture the diversity of an individual node $s$' local structure. For a node $s$ we can extract its ego network (as shown in Figure~\ref{fig_diversity}), and the diversity of $s$ is calculated as below:

\begin{equation}
    H_{t}(s) = -\sum_{j=1}^{k}p_{sj}log(p_{sj})
    \label{eq:diversity}
\end{equation}
where $k$ is the number of $s$ neighbors and $p_{sj}$ is the fraction of interactions that happen between $s$ and $j$,

\begin{equation}
    p_{sj} = \frac{w_{sj}}{\sum_{j=1}^{k} w_{sj}}
\end{equation}
where $w_{sj}$ is the weight of edge between $s$ and $j$. For the communication networks (voice calls and text messages), the Shannon entropy gives us an ego-level metric of the extent to which any one individual concentrates their communications among a few contacts or distributes them equally across contacts. For the proximity network, this measure can be interpreted as the likelihood that an individual distributes her ``face time'' equally among all of her neighbors, or spends most of her time in spatial proximity to small proportion of her total number of possible contacts. 

\subsection{Gender Diversity}
As we have seen, the study of gender segregation of friendship choice is a classic line of analysis in the literature. We extend this approach to study the extent to which men and women exhibit a preference to distribute their communications (in the case of voice calls and text messages) or ``face time'' (in the case of proximity) to same-gender or different-gender partners. To that end, we develop a \emph{gender diversity metric} (GDM) as follows. 

We can classify individuals and their neighbors as a man (m) and a woman (w). We then use the Shannon entropy to measure the gender diversity of every individual. This is given by:
\begin{equation}
    H_{g}(s) = -p_{w} log (p_{w}) - p_{m} log (p_{m})
    \label{eq:gendiv}
\end{equation}
where $p_{w}$ is the proportion of $s$' total interactions with women, and and $p_{m}$ is the proportion of $s$'s total interactions with men. For the proximity network, this measure can be interpreted as the likelihood that an individual distributes his ``face time'' equally among members of the same gender, or spends most of his time in spatial proximity to same-gender peers.

\subsection{Location Signature}
In this data, we have the traces of each student in 54 locations. For a student $s$, we denote her/his cumulative staying time in a location $L_{i}$ as $T(s, L_{i})$. In this way, for a student $s$, we can construct a vector $T_{L} (s)$,
\begin{equation}
    T_{L} (s) = \{ T(s, L_{1}), T(s, L_{2}), ..., T(s, L_{k}) \}
    \label{eq:location}
\end{equation}
where $k$ is the total number of locations. The vector $T_{L} (s)$ can be considered a signature of student $s$'s location visiting behaviors. The location similarity between two individuals $u$ and $v$ is computed by the Pearson Correlation Coefficient between the respective location vectors $T_{L} (u)$ and $T_{L} (v)$ (see Figure~\ref{fig_location}).

\subsection{Location Diversity}
\begin{figure*}[ht]
\begin{center}
\centerline{\includegraphics[width = 3.2in]{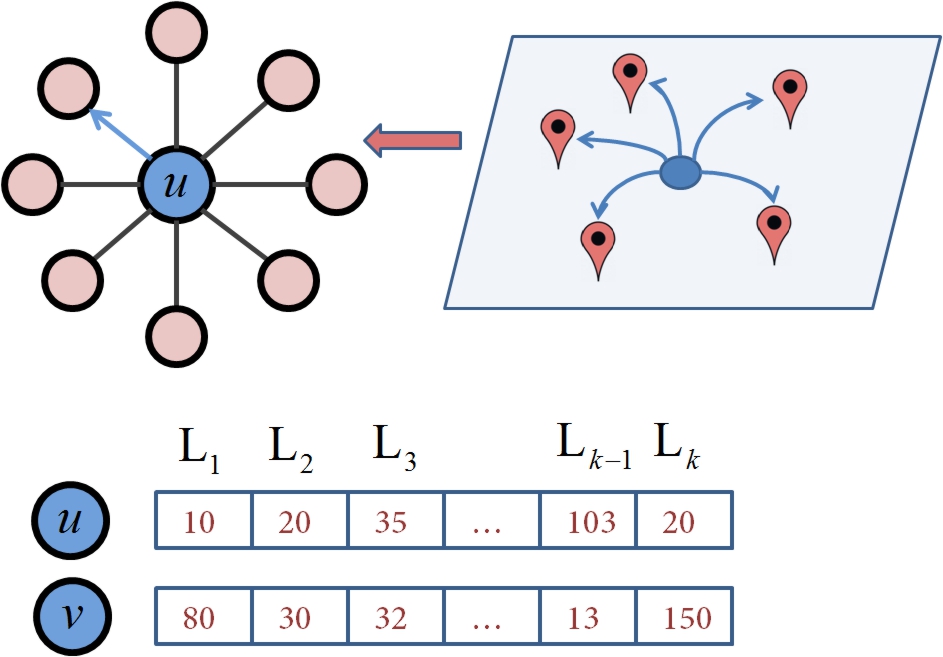}}
\caption{Location Visiting Signature Example. The ND mobility data contains traces of the time (in minutes) each individual spends in 54 campus locations. We calculate the cumulative time spent at location $L_{i}$ for each individual $u$. The resulting $1 \times 54$ vector of location durations can thus be considered that individual's spatial signature on campus.}
\label{fig_location}
\end{center}
\end{figure*}

We calculate the location diversity $H_{L}$, for each individual $s$ using that individual's location signature:
\begin{equation}
    H_{L} (s) = -\sum_{i=1}^{k} p(s, L_{i}) log(p(s, L_{i}))
    \label{eq:locdiv}
\end{equation}
where $k$ is the number of locations and $p(s, L_{i})$ is the fraction of time (hours) $s$ stays at location $L_{i}$. The location diversity $H_{L}$ captures students' preference in location visiting (see Figure~\ref{fig_location}).

\section{Gender Difference Analysis}
\label{sec_analysis}
\subsection{Significance Validation} 
We begin by ascertaining whether we observe assortative mixing by gender in each of the communication networks (voice call and text messaging). A straightforward way to measure and validate the existence of same-gender preference in communication partners is to compare the numbers of links between four possible ``mixtures'' of (directed) links (MM, WW, MW, WM) in each network  to a null model where the gender classification of node is randomly assigned, while keeping the underlying (observed) network structure intact. Positive and negative Z-scores denote the number of standard deviations by which a directed link of each type is over or under represented. We compare real data to 10,000 randomized cases where the gender classification of each node is randomly shuffled. The statistical significance of gender mixture combination is measured by the {Z} score:

\begin{equation}
    Z = \frac{N_{obs} - \mu(N_{rand})}{\sigma(N_{rand})}
    \label{eqn_gender_z}
\end{equation}
where $N_{obs}$ is the number of links by type (e.g. MM, WW, MW, WM) in the observed networks, $\mu(N_{rand})$, and $\sigma(N_{rand})$ are the average of links and the standard deviation in the gender-reshuffled networks respectively. The statistically significant Z-score values are displayed in the boldface type in the Tables. 

\begin{table}[h]
\caption{Gender-Specific Number of Calls and Text Messages Validation ($Z$-scores) ($H_{0}$ rejected at $2$ sigmas). Homophily Validation.}
\resizebox{4in}{!}{
\begin{tabular*}{\textwidth}{c @{\extracolsep{\fill}} cccccccc}
\hline \hline
&\multicolumn3c{Voice Call Network} & \multicolumn3c{Text Message Nework}\\
& Unweighted & Weighted & Weighted & Unweighted & Weighted & Weighted\\
&&(N. Events)&(Duration)&&(N. Events)&(Text Length) \\
\hline 
MM & 0.759 & 0.334 & -0.648 & -0.346 & -1.106 & -1.089 \\ 
WW & {\bf 2.164} & -0.548 & -0.431 & {\bf 2.112} & 0.331 & 0.164 \\ 
MW & {\bf -3.083} & -0.822 & 0.511 & {\bf -2.138} & 0.702 & 0.919 \\ 
WM & {\bf -2.468} & 1.022 & {\bf 1.953} & -1.649 & 0.735 & 0.735 \\
\hline
\label{tab_homo_validation}
\end{tabular*}
}
\end{table}

\subsection{Gender Differences in Assortative Mixing} 
\subsubsection{Voice Call and Text Messaging Networks}
The gender homophily analysis results for the voice call and text-messaging networks are shown in Table~\ref{tab_homo_validation}. First, it is clear that in both the voice call and text messaging networks women display statistically significant same-gender preference (both are above $2$ sigmas) but men do not ($Z$ = $0.759$ and $-0.346$ for voice call and text message networks respectively). This is consistent with previous work showing stronger same-gender preference in friendships for women in MCNs \cite{stoica2010, palchykov2012}. In both cases, cross-gender links are significantly under-represented (WM links is not significantly under-represented in the message network however, $Z$ = $-1.649$). 

In addition, when considering the overall volume of communication (the weighted representation of each network) there seems to be an over-representation of weighted cross-gender links although the z-score for this edge type is not significant ($Z$=$0.735$). Note, however, that same-gender preferences pattern are largely attenuated (dropping to non-significance) when we consider the weighted representation of each network irrespective of whether the weight represents the number of events (phone calls, messages), call duration, or text message length. This suggests that studies that rely on binary representations of friendship choices may largely overestimate the existence of gender-segregation in social networks.

\begin{table}[h]
\caption{Gender-specific proximity validation (Z-score). We examine the significance of three types of proximity interactions, MM, WW and MW/WM (proximity network is undirected, thus MW and WM are equivalent) ($H_{0}$ accepted).}
\resizebox{4in}{!}{
\begin{tabular*}{\textwidth}{c @{\extracolsep{\fill}} lcccccc}
\hline \hline
&\multicolumn2c{Semester 1} & \multicolumn2c{Semester 2} & \multicolumn2c{Semester 3} \\
 & All Pairs & Connected & All Pairs & Connected & All Pairs & Connected \\
\hline
MM & {\bf 3.645} & {\bf 3.723}  & {\bf 4.195} &  {\bf 3.906} & {\bf 1.974} & {\bf 2.044} \\
WW& {\bf 2.701} & {-0.072}  & {\bf 2.525} & -0.009 & {\bf 2.638} & {0.583} \\
MW/WM & {\bf -7.907} & {\bf -4.444}  & {\bf -8.745} & {\bf -4.466}  & {\bf -5.616} & {\bf -3.652} \\
\hline
\end{tabular*}
}
\label{tab_proximity_homo}
\end{table}

\subsubsection{Proximity Network}
Table~\ref{tab_proximity_homo} shows the corresponding gender homophily analysis results for the proximity network. We partition the proximity log into three temporal snapshots. The first one is from Aug of 2011 to Dec of 2011, corresponding to the first semester that students spent on campus (Semester 1); the second snapshot goes from Jan of 2012 to May of 2012 (Semester 2); and the third snapshot covers the period from Aug of 2012 to Dec of 2012 (Semester 3). We do not consider proximity data from other time periods, since these are logged during summer and winter vacations when the majority of students are off campus. Additionally when calculating the cumulative proximity time between two students (how long they have been proximate to each other in one semester), we exclude proximity times logged between 9:00 PM and 9:00 AM, in order to remove large (spurious) proximity values generated by students who spend time sleeping in the same dorm. For each proximity metric, we present separate sets of results for all of the participants in the study (All Pairs) and for those participants who have an edge in either the voice call or text messaging networks (Connected).

The results of our analysis show that in contrast to the communication-based networks, men exhibit a stronger levels of assortative mixing by gender than women. The all-male (MM) dyad is overrepresented in the proximity network in all three time periods ($Z > 1.96$), whether we consider all pairs or only respondents who are directly connected via a communication link. Women on the other hand, exhibit a tendency to be spatially proximate to other women (WW dyad) when we consider all pairs, but this tendency disappears when we restrict our analysis to those individuals with whom they are connected. This suggests that while both men and women experience high levels of spatially based gender assortativity, only men seem to couple direct social connectivity with spatial contiguity. Women on the other hand seem to spend equal amounts of time being spatially proximate to their social contacts regardless of their gender. Finally, mixed gender dyads are significantly underepresented in all three periods for both all pairs and connected dyads ($Z < -2$). 

\begin{table}[h]
\caption{Gender-Specific Number of Calls and Text Messages Validation ($Z$-scores) ($H_{0}$ rejected at $2$ sigmas): Action Validation.}
\centering
\resizebox{4in}{!}{
\begin{tabular*}{\textwidth}{c @{\extracolsep{\fill}} cccccccc}
\hline \hline
&\multicolumn3c{Voice Call Network} & \multicolumn3c{Text Message Nework}\\
& Unweighted & Weighted & Weighted & Unweighted & Weighted & Weighted \\
& &(N. Events)&(Duration)& &(N. Events)&(Text Length) \\
 \hline 
From M & -0.629 & -0.187 & -0.428 & {\bf -1.971} & -0.920 & -0.730\\ 
From W & 0.629 & 0.187 & 0.428 & {\bf 1.971} & 0.920 & 0.730\\ 
\hline
To M & -0.544 & 1.159 & 0.082 & -1.152 & -0.897 & -0.862 \\ 
To W & 0.544 & -1.159 & -0.082 & 1.152 & 0.897 & 0.862 \\ 
\hline
\label{tab_action_validation}
\end{tabular*}
}
\end{table}

\subsection{Gender Differences in Sociability (Outgoing Communications)} 
A long standing finding in the MCNs (mobile communication networks) and SNSs (social network sites) literature is that women tend to initiate and produce a larger volume of interactions than men \cite{lewis2008, miritello2013}. As shown in the first two rows of the bottom panel of Table~\ref{tab_action_validation}, we find mixed support for this hypothesis. In the phone network men and women are equally likely to be the initiator of communications. In the message network, we observe that women are indeed more likely to initiate communications ($Z$ = $1.971$). This pattern of results suggests that dispositional gender differences in network-related behaviors may depend on the communication channel that is being considered. Below we consider other results that are consistent with this expectation. 

\subsection{Gender Differences Popularity (Incoming Communications)}
Previous work in online environments (i.e., social network sites) has shown that women are more likely to be the target of communication attempts \cite{griffiths2004}. As shown by the last two rows of Table~\ref{tab_action_validation} we find little support for this hypothesis in our data ($|Z| < 2$). Men and women are equally likely to be the recipient of communications in both the voice call and text message networks. This result remains the same whether we consider the binary or weighted representation of each network. 

\begin{table}[h]
\caption{Gender-specific Reciprocity of Call and Text Message Validation ($Z$-scores).}
\centering
\resizebox{4in}{!}{
\begin{tabular*}{\textwidth}{c @{\extracolsep{\fill}} ccc}
\hline
& {Voice Call Network} & {Text Message Network} \\ \hline \hline
{MM}& {\bf 2.019} & {\bf -1.969} \\
{WW}& -0.312 & -1.005 \\
{MW}& -0.501 & {\bf 2.541} \\
{WM}& -1.515 & 1.171 
\\
\hline
\label{tab_reciprocity}
\end{tabular*}
}
\end{table}

\subsection{Differences in Reciprocity by Gender Mix}
We consider the question of whether directed links are more likely to be reciprocated depending on whether these links are same-gender versus cross-gender links and whether the participants in the interaction are men or women. To that end, we follow the same procedure used above to examine homophily patterns. First, we compute the observed rate of reciprocation along the four types of links classified by the gender mix. Second, we compare the observed frequencies with those obtained from a null model where the gender classification of each node is randomly assigned but the network structure remains intact. In each round of simulation, we obtain counts of reciprocated relationships for the four link types in the randomized network. The simulation is repeated for 10,000 times. Finally we use equation~\ref{eqn_gender_z} to compute a reciprocation z-score for each gender mixture.

Table~\ref{tab_reciprocity} shows the results. We find that relations between men (MM dyads) are significantly more likely to be reciprocated in the voice call network than we would expect by chance. However, we find the \emph{opposite} pattern of results in the text message network; here men are less likely to reciprocate interactions initiated by other men. In addition, we find significant differences in reciprocation behavior for cross-gender pairs depending on the gender of the initiator. Women are more likely to reciprocate a text message coming from a man than we would expect by chance ($Z = 2.541$), but men are equally likely to reciprocate a text message regardless of the sender's gender ($Z = 1.171$). Women exhibit no tendencies for higher levels of within-gender reciprocity than we would expect by chance on either the voice call or text messaging network ($|Z| < 2$).

\subsection{Gender Differences in Topological and Gender Diversity}
\subsubsection{Voice Call and Text Messaging Networks}
We define the topological diversity as an ego-level metric capturing an individual's tendency to concentrate communication on a few contacts or distribute them evenly across neighbors (see equation~\ref{eq:diversity}). We define the gender diversity as ego-level metric capturing the tendency to concentrate communications either men or women contacts or to spread communications evenly across gender lines (see equation~\ref{eq:gendiv}). Both measures range from 0 to 1, with a higher value indicating a greater propensity to distribute communications equitably across neighbors or gender categories. A key question is whether men and women differ in both of these behavioral propensities. As before we generate a null model by keeping the distribution of link weights the same but reshuffling the gender labels and calculating the topological and gender diversity for each individual at each iteration. We compute Z-scores using the procedure outlined above. 

\begin{table}[h]
\caption{Gender-Specific Diversities Validation (Z-scores). $H_{0}$ is rejected at 2 sigmas.}
\resizebox{4in}{!}{
\begin{tabular*}{\textwidth}{c @{\extracolsep{\fill}} lcccc}
\hline \hline
& \multicolumn2c{Voice Call Network} & \multicolumn2c{Text Message Network} \\  
& $H_{g}$ & $H_{t}$ & $H_{g}$ & $H_{t}$ \\ \hline
Gender Effect (Women) & {\bf 2.148} & {\bf 2.041} & {\bf 2.913} & {\bf -2.006} \\ \hline
\label{tab_z_score_topological}
\end{tabular*}
}
\end{table}

Table~\ref{tab_z_score_topological} shows the results. We find that women differ systematically from men in their tendency to concentrate communications on a few contacts but that the direction of this difference depends on the communication channel (voice versus text). In the voice call network, women are more likely than men to distribute their communications evenly across contacts \emph{and} gender categories. In the text messaging network, on the other hand, women are more likely than men to concentrate their social interactions on a few persons, but are \emph{less} likely to concentrate their communications on members of only one gender. This suggests that women tend to be more likely to structure their communication messaging networks in terms of a core/periphery regime, but that the \emph{gender} of the preferred contact is not necessarily predictable. Men on the other hand are more likely to partition their voice call network in terms of a core-periphery regime, \emph{and} they tend to select a same-gender alter as a preferred voice call contact. 

\begin{table}[h]
\caption{Gender Diversity and Topological Diversity in Proximity Network (Z-score). We observe that women have significantly higher gender diversity and topological diversity than men. This observation persists in three semesters. ($H_{0}$ rejected at 2 sigmas)}
\resizebox{4in}{!}{
\begin{tabular*}{\textwidth}{c @{\extracolsep{\fill}} ccccccc}
\hline \hline
&\multicolumn2c{Semester 1} & \multicolumn2c{Semester 2} & \multicolumn2c{Semester 3} \\
& $H_{g}$ & $H_{t}$ & $H_{g}$ & $H_{t}$ & $H_{g}$ & $H_{t}$  \\
\hline
Gender Effect (Women)& {\bf 1.972} & {\bf 1.960}  & {\bf 2.067} &  {\bf 2.077} & {\bf 1.960} & {\bf 1.966} \\
\hline
\end{tabular*}
}
\label{tab_proximity_diversity}
\end{table}

\subsubsection{Proximity Network}
Do men and women exhibit similar patterns of topological and gender diversity when it comes to spatial proximity as they do for voice calls and text messages? Table~\ref{tab_proximity_diversity} shows the results. We find that women have higher topological and higher gender diversity than men throughout all three periods. This suggests that, in contrast to men, who seem to segment their face-to-face interactions so that they end up spending a disproportionate amount of time spatial proximity to a select group of other men, women distribute their ``face time'' equally across all of their possible contacts regardless of their gender category. 

\subsection{Gender Differences in Communication and Proximity Coupling}
It is unlikely that the gender-based dynamics in the communication (voice call and text message) networks and the proximity networks are independent. Instead, social network theory suggests that there are fundamental dependencies between the two, such that propinquity leads to connectivity via communicative interactions, and communicative interactions lead to propinquity by increasing the motivation and opportunities to engage in face to face interaction \cite{feld1981, carley1991, doreian2012}. However, our results suggest that there should exist systematic differences between men and women in the extent to which social connectivity affects subsequent patterns of spatial proximity and face-to-face interaction. In essence, men seem to couple social interaction and physical proximity in a stronger way than women (see the results summarized in Table~\ref{tab_proximity_diversity}). We evaluate this hypothesis by ascertaining whether we can observe significant changes in the proximity network once we know that two individuals are connected in the communication network and whether this effect is the same for men and women. 

\subsubsection{Procedure}
We proceed as follows. First, we calculate the daily average proximity ($t$) for each pair of individuals, $u$ and $v$, before they are connected in the communication network, we then calculate the same quantity ($t'$) \emph{after} we observe those individuals interact via voice call or text messaging. We then compute the difference between these two quantities ($\Delta t$ = $t'-t$). This gives us an estimate of the effect of connectivity on time spent in social proximity. We then use the same null model generating procedure (see Equation~\ref{eqn_gender_z}) described above to determine the existence of significant differences in proximity differences between pre and post-connectivity proximities for same-gender (MM and WW) and mixed gender (MW/WM) dyads.

\begin{figure}[ht]
	\centering
	   \includegraphics[width=2.5in]{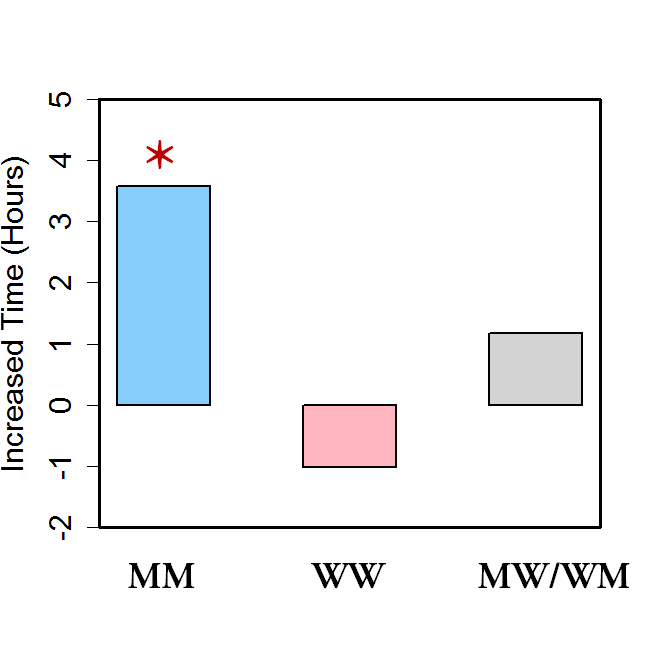}
	\caption{The impact of friendship on mobility behaviors. All-men dyads (MM) display higher levels of spatial proximity after connectivity in the Voice Call and Text Messaging networks. This effect is not observed in the all-women (WW) and the mixed-gender (WM/MW) dyads}.
	    \label{fig_beforeafter_proximity}
\end{figure}

\subsubsection{Results}
The results are summarized in Figure~\ref{fig_beforeafter_proximity}. We find that, consistent with expectations, spatial proximity increases after observing connectivity in the communication network only for the all-men (MM) dyad. After being connected in the communication network, members of the average all-male dyad are observed to spend about 3.5 additional hours near one another, which is a substantively significant effect. For all-women (WW) and the mixed gender (MW/WM) dyad there is no statistically appreciable effect of connectivity on subsequent patterns of spatial proximity. This supports the hypothesis that men tend to couple social and physical connectivity, such that they end up spending more time in gender-segregated groups than women do.

\begin{table}[h]
\caption{Gender-specific difference in location diversity (Z-score). We observe that women have significantly higher location diversity than men. This observation persists in three semesters. ($H_{0}$ rejected at 2 sigmas)}
\resizebox{4in}{!}{
\begin{tabular*}{\textwidth}{c @{\extracolsep{\fill}} lcccc}
\hline \hline
&Semester 1 & Semester 2 & Semester 3 \\
\hline
Men& {\bf -2.183} & {\bf -1.958}  & {\bf -2.250} \\
Women& {0.052} & {0.031}  & {-0.057} \\
\hline
\end{tabular*}
}
\label{tab_location_diversity}
\end{table}

\subsection{Gender Differences in Mobility Behaviors}
\label{sec_location}
\subsubsection{Location Diversity}
In this section we address the question of whether we can observe any gender differences in location diversity patterns. Table~\ref{tab_location_diversity} shows the results. We find that on the whole women display higher levels of location diversity than men. This mirrors in terms of spatial mobility behavior the way that women distribute their communication and face-to-face interaction time across alters: in both of these respects women are less ``predictable'' than men. Note that this result has the implication that men not only spend more time interacting in gender-segregated groups with a relatively smaller number range of individuals, but that they may do so in a smaller set of geographic locations than women, including being more likely to use the same-gender segregated spatial settings repeatedly. In this respect, there seems to be an empirical connection between the male pattern of gender-segregated spatial interaction and their relatively more predictable patterns of spatial mobility. 

\begin{figure*}[ht]
	\centering
	\centerline{\includegraphics[width=4in]{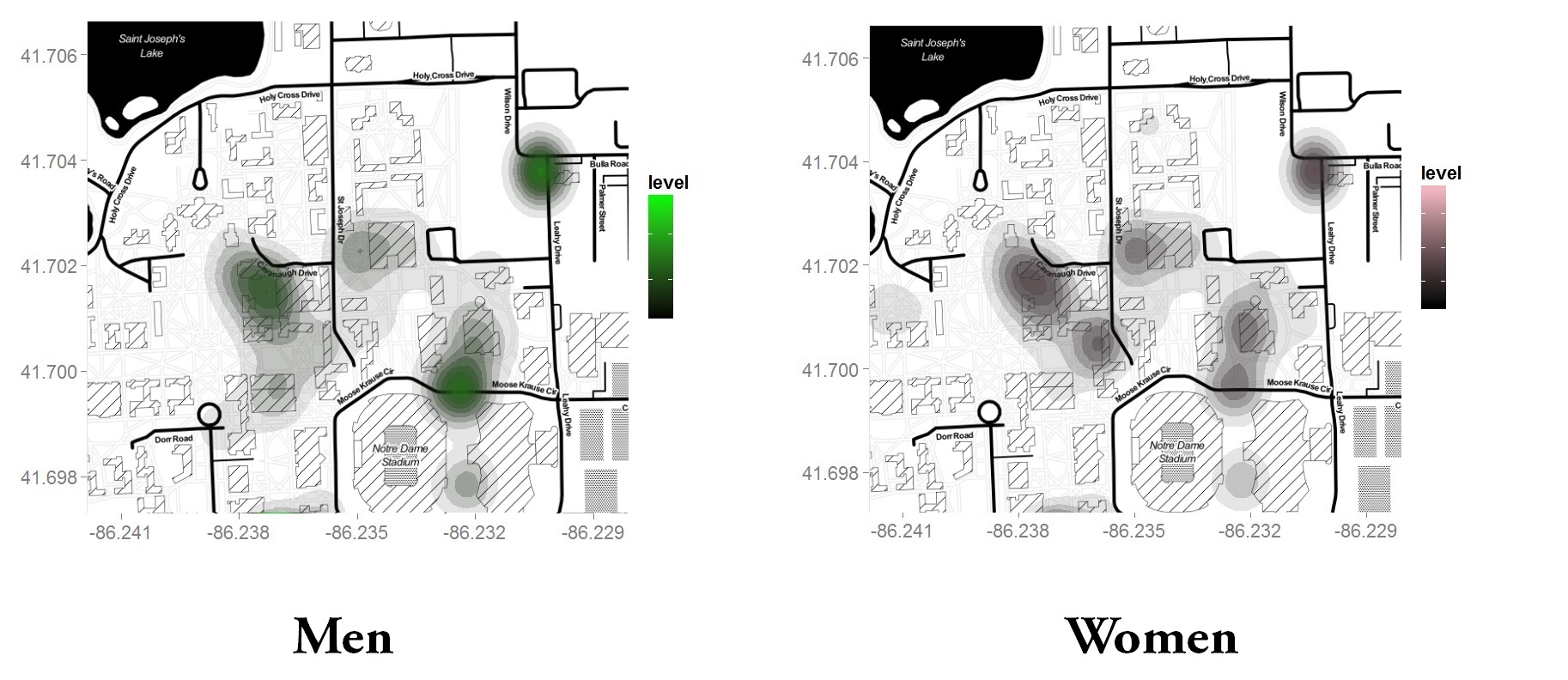}}
	\caption{Location Stay Density of Women and Men. Green indicates men, and pink indicates women. The higher brightness indicates larger concentration density.}
\label{fig_gender_density}
\end{figure*}

Figure~\ref{fig_gender_density} provides a visual representation of this effect, overlaid against an actual representation of the campus spatial layout. The figure provides clear evidence of the relative concentration of spatial mobility for men and the relative dispersion of mobility for women. On the whole men are more likely to be found frequenting the same locations on campus for longer periods of time. Women on the other hand disperse their activities across a wider range of locations, leading them to spend relatively smaller amounts of time at any one location. 

\begin{table}[h]
\caption{Gender-specific location visiting validation (Z-score). We examine the significance of three types of location visiting similarity, MM, WW and MW/WM.}
\resizebox{4in}{!}{
\begin{tabular*}{\textwidth}{c @{\extracolsep{\fill}} lccccc}
\hline \hline
&\multicolumn2c{Semester 1} & \multicolumn2c{Semester 2} & \multicolumn2c{Semester 3} \\
Types of Pairs\\
& All Pairs & Connected & All Pairs & Connected & All Pairs & Connected \\
\hline
MM & {\bf 4.136} & {\bf 4.718}  & {\bf 3.913} &  {\bf 4.108} & {\bf 3.077} & {\bf 3.450} \\
WW& {0.548} & {1.701}  & {-1.208} & 0.590 & {-1.731} & {-0.097} \\
MW/WM & {\bf -10.229} & {\bf -8.730}  & {\bf -9.096} & {\bf -5.825}  & {\bf -6.054} & {\bf -5.345} \\
\hline
\end{tabular*}
}
\label{tab_location_homo}
\end{table}

\subsubsection{Similarity in Location Signature by Dyadic Gender Mix}
Are same-gender dyads more likely to share mobility habits than cross gender dyads? To answer this question, we develop a measure of dyadic similarity in location signatures by computing the Pearson correlation coefficient $r_{uv}$ between the location signature vectors of each individual at each time period. We then use the null model validation method described above to ascertain whether this correlation is higher than one would expect by chance across the same gender (MM, WW) and across different genders (MW/WM) dyads. We generate validation results both for all pair of individuals (All Pairs) and for individuals who are connected in the Voice Mail or the Text Message communication networks (Connected). Consistent with the results reported above, we expect to find higher levels of location signature similarities among all-men dyads (MM) than in the other gender-based dyadic combinations. 

The results are shown on Table~\ref{tab_location_homo}. In support of our male similarity hypothesis, we find that members of all-men dyads tend to display higher levels of location signature similarity than we would expect by chance. This location signature similarity pattern emerges in the first semester, and can be found in all subsequent time periods. For the all-women dyad, location signature similarities are close to what we would expect under a random-pairing regime. For cross-gender dyads, on the other hand, we find a smaller correlation between location signature pattern than predicted under a random pairing regime. 

This pattern of results is the same irrespective of whether we compare all pairs of individuals or only those directly connected in the communication network. This suggests that the location signature correlation is independent of social connectivity (it is not the result of a direct social influence process) but may reflect common underlying dispositional attributes correlated with both the odds of individuals forming a social connection and their underlying mobility habits. These results reinforce the impression that men tend to have more predictable and have less diverse mobility habits than women. Men also tend to couple mobility and proximity behaviors in a strong way. In addition, the fact that men and women tend to have such distinct location signature patterns, means that any cross-gender pair drawn at random from the population is likely to have negatively correlated location signature patterns, whether they share a direct social link or not. 

\section{Discussion}
\label{sec_discussion}
In this paper we have examined gender differences in dispositional factors relevant for the formation and maintenance of social networks. We attempted to isolate dispositional differences between men and women by studying a population in a relatively closed context at the same stage of the life course (the transition to college) before the onset of life-transitions that tend to exacerbate structurally induced (non-dispositional) differences in the social connectivity behavior between men and women (e.g. women's greater reliance on kin; men's greater likelihood to connect to co-workers). We also extended previous work by looking at patterns of behavior (e.g. distributional and proximity tendencies) where gender differences have yet to be documented systematically. 

Building on previous work that has looked at gender differences in social networks with unobtrusively collected data, we have examined gender differences across both communication (cell phone and text messaging) and mobility (spatial proximity) networks. Our results are consistent with (and thus validate in a natural setting) previous work done on MCNs and SNSs \cite{lewis2008, kovanen2013, stoica2010, dong2014, palchykov2012}. They are also consistent with an emerging line of research that suggests that gender differences in behavioral dispositions associated with mobility and spatial proximity behaviors differ in significant ways than those associated with communication behavior \cite{fournet2014, stehle2013}. The key implication of our study therefore is that when studying gender differences in social networks, researchers should be wary of extending conclusions reached when studying one realm of social behavior (e.g. communication) into an unrelated realm (spatial proximity patterns). 

For instance, our analysis of same-gender preference in both (voice call and SMS) of our communication networks turned up the now well-documented pattern of stronger same-gender preference for women, but the gender differences were more pronounced in the binary representation of the network than when call volume was taken into account. Most importantly, however, we found the \emph{opposite} same-gender preference pattern in the proximity network, with men displaying stronger behavioral patterns of same-gender proximity than women. These patterns emerged early on (after the first semester) and remained over a long temporal scale (one year). Gender differences in spatial proximity behavior also extended to the \emph{distribution} of face-to-face interaction across contacts. In all, men were more likely to unevenly distribute their face time in favor of same-gender alters, and were more likely to concentrate their face-to-face time on a single contact. This contrasts with the results obtained in the communication (SMS) network, which showed that women were more likely to concentrate their communication behavior on a few contacts than men. 

Related to the tendency of men to concentrate their face-to-face interaction on same gender contacts, we found that there is a stronger linkage between communication and subsequent time spent in close proximity for men but not for women, suggesting that men tend to couple their face-to-face time to with their previous communicative interactions in a stronger way than women. The tendency of men to form gender-segregated ``bands'' that spend a lot of time in close proximity affect the predictability of their mobility behavior: On the whole, we found that men's mobility behavior (as given by the time spent across a set of demarcated locations) is more predictable than that of women because it tends to concentrate itself in a smaller number of locations. In accordance with this result our analysis reveals that there is a stronger level of concordance in spatial mobility habits in all-male dyads than there is in mixed-gender or all-women dyads. 

In all the results reported in this paper paint a picture of substantial differences in the mobility and spatial proximity behavior of men and women. These differences stand in stark contrast to the relatively small gender-differences found when looking at purely communicative interactions, supporting the view that communication and exchange networks should not be analyzed in isolation from their embedding in a spatial context \cite{doreian2012}. 

The greater likelihood of men's proximity and spatial behavior, and the greater-similarity of all-male dyads in mobility habits is consistent with previous sociological work. These studies note that men's friendship same and cross-gender selection and behavior is driven by more by dispositional factors than those of women, who seem be more likely to respond to the opportunity structure of contacts \cite{kalmijn2002}. Our results validate a long line of ethnographic observational research showing the stronger effect of same-gender peer groups constructed around such activities as games and sports on the ability of men to construct coherent masculine identities especially during key transition points in the life course \cite{messner1990, maccoby1988, smith1993}. Our results are also consistent with social psychological perspectives that point to the strong affinity of males, beginning at young ages, to interact in relatively larger groups in specific settings outside the household while strongly avoiding cross-gender interactions in the same settings \cite{ridgeway1999, mehta2009}. 

Our finding of greater spatial homophily among men is in line with dispositional accounts of gender differences in friendship patterns at the intersection of the social and biological sciences (e.g. evolutionary psychological frameworks). This line of work is beginning to move beyond traditional scholarly understandings of women as predisposed to be more ``social'' than men \cite{cross1997}, and towards a more comprehensive account that sees men and women as predisposed to distinct forms of sociality and intimacy \cite{baumeister1997, benenson2009}. These researchers point to the ubiquity and functional role (e.g. mutual protection, resource hoarding, etc.) of stronger dispositions towards homosociality among men across cultures and historical settings \cite{benenson2014}. This is evidenced by the greater likelihood of observing ``coalitions'' of unrelated males in close physical proximity especially in public spaces \cite{rodseth2000, vigil2007}; the greater investment (e.g. reported enjoyment and interaction preferences) displayed by males (starting early in life) same-gender peer relationships centered in specific spatial contexts, such as neighborhoods and classrooms \cite{benenson1998}; the greater likelihood that same-gender relationships among men of being embedded in a larger group context \cite{benenson2003}; women's greater preference for interaction in dyads rather than groups \cite{benenson1997}; the greater dispositional capacity of males to resolve conflict with same-gender peers in comparison to females \cite{benenson2014}; and greater likelihood of same-gender best-friendships among males to endure longer and to survive in the face of relationship conflict and strain \cite{benenson2009} .

The basic empirical implication of these perspectives is that we should expect to find predictable \emph{asymmetries} in the way that men and women structure the temporal and spatial organization of their same-gender peer relations \cite{vigil2007}. While men prefer socialize in relatively large, function-specific groups that interact in public settings, women prefer to interact across emotionally intimate dyads socializing in private settings. The asymmetry hypothesis reconciles the discrepant findings of stronger same-gender preference and greater unevenness in the distribution of communications tendencies in the SMS network for women and the stronger same-gender preference and greater unevenness distribution of spatial interactions in the proximity network for men. In all, our results provide striking support for the proposition that there is an intimate connection between social and communicative behavior and mobility and proximity patterns but that the empirical linkage between these is governed by cultural, personality, and possibly biological factors (glossed under the dispositional view) that differ systematically between men and women.

\bibliographystyle{nws}
\bibliography{gender}

\end{document}